\documentclass[sigconf]{acmart}
\pdfoutput=1

\renewcommand\footnotetextcopyrightpermission[1]{}

\usepackage[]{hyperref}
\usepackage[capitalize]{cleveref}

\usepackage{braket}
\usepackage{amsmath}

\usepackage{amssymb}

\usepackage[ruled]{algorithm2e}

\usepackage{multirow}

\usepackage{color}

\usepackage{adjustbox}

\usepackage{listings}

\usepackage[english]{babel}

\usepackage{capt-of}

\usepackage{dblfloatfix}

\usepackage{graphicx}

\definecolor{orange}{RGB}{255,165,0}

\lstdefinelanguage{QuMIS}
{
  morekeywords={
    Pulse,
    MPG,
    MD,
    wait
  },
  morekeywords=[2]{},
  morekeywords=[3]{mov,add,addi,sub,nop,beq,bne, QNopReg},
  morekeywords=[4]{Outer\_Loop},
  morekeywords=[5]{I, X180, Y180, Ym90, Y90, CZ},
  sensitive=false, 
  morecomment=[l]{\#}, 
}

\definecolor{mygray}{gray}{0.6}
\newcommand{\gray}[1]{\textcolor{mygray}{\footnotesize{#1}}}

\newcommand{\ns}{\mathrm{ns}}
\newcommand{\us}{\mu\mathrm{s}}
\newcommand{\GHz}{\mathrm{GHz}}
\newcommand{\MHz}{\mathrm{MHz}}
\newcommand{\fq}{f_\mathrm{Q}}
\newcommand{\fr}{f_\mathrm{R}}
\newcommand{\Tone}{T_1}
\newcommand{\portf}{P_{\mathrm{F}}}
\newcommand{\porto}{P_{\mathrm{o}}}
\newcommand{\porti}{P_{\mathrm{i}}}

\title{An Experimental Microarchitecture for a Superconducting Quantum Processor}
\author{X.~Fu\textsuperscript{1,2,$\ast$} ~~ M.~A.~Rol\textsuperscript{1,3} ~~ C.~C.~Bultink\textsuperscript{1,3} ~~ J.~van~Someren\textsuperscript{1,2} ~~ N.~Khammassi\textsuperscript{1,2} ~~ I.~Ashraf\textsuperscript{1,2} R.~F.~L.~Vermeulen\textsuperscript{1,3} ~~ J.~C.~de~Sterke\textsuperscript{4,1} ~~ W.~J.~Vlothuizen\textsuperscript{5,1} ~~ R.~N.~Schouten\textsuperscript{1,3} ~~ C.~G.~Almudever\textsuperscript{1, 2} ~~ L.~DiCarlo\textsuperscript{1,3,$\dagger$} ~~ K.~Bertels\textsuperscript{1,2,$\ddagger$}}
\affiliation{
  \institution{\textsuperscript{1} QuTech, Delft University of Technology, P.O. Box 5046, 2600 GA Delft, The Netherlands\\
\textsuperscript{2} Computer Engineering Lab, Delft University of Technology\\
\textsuperscript{3}Kavli Institute of Nanoscience, Delft University of Technology\\
\textsuperscript{4} Topic Embedded Systems B.V.\\
\textsuperscript{5} Netherlands Organisation for Applied Scientic Research (TNO)}
\{$\ast$~x.fu-1,~~$\dagger$~l.dicarlo,~~$\ddagger$~k.l.m.bertels\}@tudelft.nl
}

\date{\today}
\begin{document}

\begin{abstract}
Quantum computers promise to solve certain problems that are intractable for classical computers, such as factoring large numbers and simulating quantum systems. To date, research in quantum computer engineering has focused primarily at opposite ends of the required system stack: devising high-level programming languages and compilers to describe and optimize quantum algorithms, and building reliable low-level quantum hardware. Relatively little attention has been given to using the compiler output to fully control the operations on experimental quantum processors. Bridging this gap, we propose and build a prototype of a flexible control microarchitecture supporting quantum-classical mixed code for a superconducting quantum processor. The microarchitecture is based on three core elements: (i) a codeword-based event control scheme, (ii) queue-based precise event timing control, and (iii) a flexible multilevel instruction decoding mechanism for control. We design a set of quantum microinstructions that allows flexible control of quantum operations with precise timing. We demonstrate the microarchitecture and microinstruction set by performing a standard gate-characterization experiment on a transmon qubit.
\end{abstract}
\maketitle

\section{Introduction}
\begin{figure}[t]
\centering \includegraphics[width=0.9\linewidth]{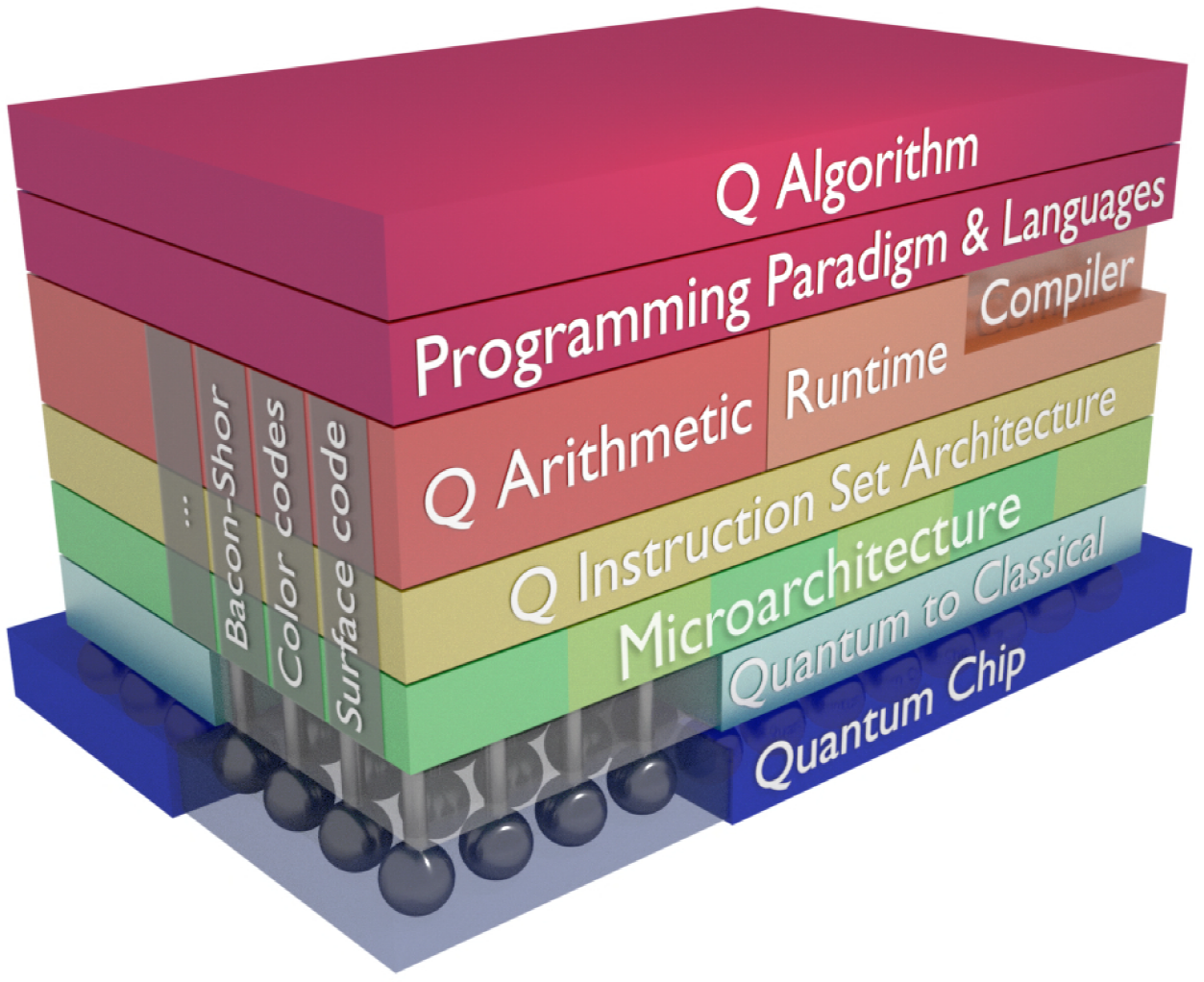}
\caption{Overview of the quantum computer system stack from~\cite{fu2016heterogeneous}.}
\label{fig:stack}
\end{figure}

To construct a fully programmable quantum computer based on the circuit model~\cite{nielsen2010quantum}, a system stack~\cite{fu2016heterogeneous} composed of several layers is required (Figure~\ref{fig:stack}). Quantum algorithms are formulated and then described using a high-level quantum programming language~\cite{omer2003structured, abhari2012scaffold, green2013, wecker2014liqui, Steiger2016projectq}.
Depending on the choice of quantum error correction code~\cite{terhal2015quantum}, such as surface code~\cite{fowler2012surface}, the compiler~\cite{svore2006layered, wecker2014liqui, javadiabhari2015scaffcc} takes that description as input, performs optimization~\cite{amy2016verified, kudrow2013quantum, paetznick2014repeat, wecker2014liqui, heckey2015compiler} and generates a fault-tolerant implementation of the original quantum algorithm. Next, it realizes the algorithm using instructions~\cite{balensiefer2005evaluation, svore2006layered, javadiabhari2015scaffcc, smith2016practical, ibmqasm20} belonging to a quantum instruction set architecture (QISA). Just like in classical architectures~\cite{Hennessy2011computer}, the QISA is the interface between software and hardware. A control microarchitecture is needed to decode the quantum instructions into required control signals with precise timing as well as real-time quantum error detection and correction~\cite{dennis2002topological, fowler2015minimum}.
Finally, based on the specific quantum technology -- e.g., superconducting qubits~\cite{kelly2015state, riste2015detecting, kandala2017hardware}, trapped ions~\cite{monroe1995demonstration, debnath2016demonstration}, spin qubits~\cite{Hanson2007Spins}, nitrogen-vacancy centers~\cite{de2010universal, cramer2016repeated}, etc. -- control signals are translated into required pulses, and sent to the quantum chip via the quantum-classical interface.

In current experiments, quantum processors are controlled with well-defined electrical signals, e.g., microwave-frequency and baseband pulses, which require accurate parameters and timing. To satisfy the strict requirements on control signals, dedicated electronic devices are typically used to interface with the quantum processor. However, existing control methods introduce high resource consumption, long configuration times, and control complexity, all of which scale poorly with the number of qubits~\cite{Hornibrook2015Cryogenic}. Although high-level languages offer flexibility, quantum compilers typically generate instructions that are not directly executable on a quantum processor. It is a challenge to design a control microarchitecture that accepts a set of instructions output by a compiler and translates them into the interface required by a quantum processor.

Motivated by heterogeneous computing, we propose a control microarchitecture, named QuMA, for a superconducting quantum processor based on the circuit model. QuMA accepts quantum-classical mixed code and enables flexible and precise-timing control over a quantum processor. The four concepts at the core of QuMA are:
\begin{itemize}
\setlength\itemsep{0pt}
    \item Codeword-based event control scheme: every event including pulse generation and measurement is assigned with an index, which is called a codeword. These events are triggered by corresponding codewords at runtime. This scheme abstracts the control of quantum processors using complex analog pulses into a simple interface consisting of only handy binary signals, providing the foundation for flexible control via instructions.
    \item Queue-based event timing control: in this scheme, events with precise timing decoded from instruction execution are first buffered in a group of queues and then triggered at expected timing. It allows that events are triggered at deterministic and precise timing while the instructions are executed with non-deterministic timing.
    \item Multilevel instruction decoding: quantum instructions are successively translated into  microinstructions, micro-opera-tions, and finally codewords with accurate timing. It enables using technology-independent instructions to control operations on qubits.
    \item Quantum microinstruction set: we design and implement a low-level quantum microinstruction set (QuMIS) which enables flexible control of quantum operations.
 \end{itemize}
In addition, we implement QuMA on a field-programmable gate array (FPGA). We experimentally validate QuMA by conducting a standard gate-characterization experiment on a superconducting qubit, which is called \textit{AllXY}~\cite{chow2010optimized, reed2013entanglement}. The control, initially specified in a high-level programming language, is converted to our proposed instructions by a quantum compiler.

The paper is structured as follows. Section~\ref{sec:bgd} briefly introduces the basics of quantum computing and the superconducting qubits as used in the experiment. Section~\ref{sec:rr} presents related previous work. After stating the challenges of controlling quantum processors using instructions in Section~\ref{sec:challenge}, Section~\ref{sec:uarch} details how QuMA addresses these challenges in a systematic way with three proposed mechanisms. Section~\ref{sec:scale} discusses the advantages and scalability of QuMA. The implementation and experimental validation of QuMA and QuMIS are shown in Sections~\ref{sec:implementation} and~\ref{sec:exp}, respectively. Section~\ref{sec:conc} concludes.
\section{Background}
\label{sec:bgd}
\subsection{Quantum Computing Basics}
Quantum computing can be best viewed as computation-in-memory, in which information is stored and processed at the same place with the basic elements called qubits.
A qubit can exist in a superposition of its two logical states, $\ket{0}$ and $\ket{1}$, which is mathematically described by $\ket{\psi} = \alpha\Ket{0} + \beta\Ket{1}$, where $\alpha,\beta\in \mathbb{C}$ satisfy $\lvert\alpha \rvert ^2 + \lvert \beta \rvert ^2=1$. The state of a qubit can be intuitively depicted by a vector on the Bloch sphere~\cite{nielsen2010quantum}. When measured in the logical basis, a qubit is projected onto  $\Ket{0}$ or $\Ket{1}$ with probabilities  $\lvert\alpha\rvert^2$ and $\lvert\beta\rvert^2$, respectively.

The qubit state can be modified by applying quantum gates. Every single-qubit gate is a rotation $R_{\hat{n}}(\theta)$ on the Bloch sphere along an particular axis $\hat{n}$ by an angle $\theta$. Popular single-qubit gates include $R_x(\pi)$, $R_y(\pi)$, and $R_z(\pi)$, which are also called $X$, $Y$, and $Z$, respectively.
There are also two-qubit gates, among which the most popular are the controlled-NOT (CNOT) and the controlled-phase (CZ). 
For a comprehensive introduction to quantum computing basics, we refer the interested reader to~\cite{nielsen2010quantum}.

\subsection{Superconducting Qubits}
\label{sec:transmon}
In this paper, we focus on transmon qubits~\cite{koch2007charge} in planar circuit quantum electrodynamics~\cite{blais2004cavity}. This is a promising architecture for solid-state quantum computing where qubit measurement and a universal gate set~\cite{divincenzo2000physical}, comprised of single-qubit gates (mainly $X$ and $Y$ rotations) and the CZ gate, have already achieved error rates lower than the fault-tolerance threshold for surface code~\cite{fowler2012surface}. 
Recent experiments have demonstrated basic quantum error correction for this architecture, including the repetition code~\cite{riste2015detecting, kelly2015state} and elements of the surface code~\cite{takita2016demonstration}. 

\begin{figure}[hbt]
\centering \includegraphics[width=\linewidth]{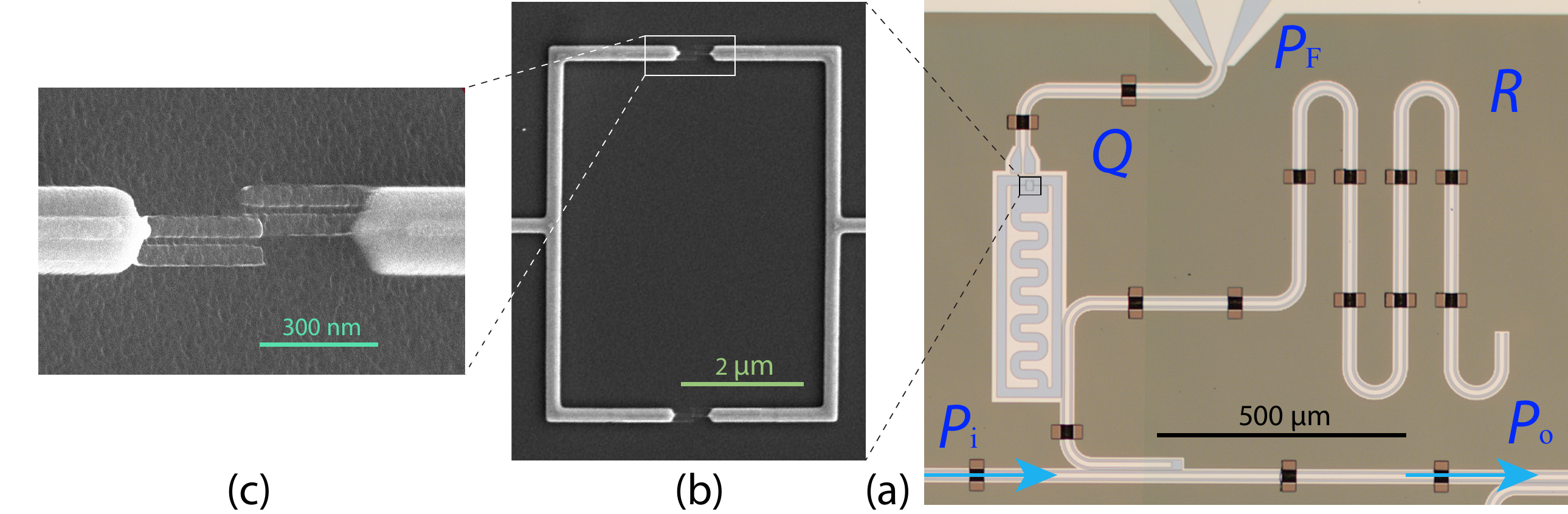}
\caption{Images at various scales of a transmon qubit coupled to a readout resonator in a planar circuit quantum electrodynamics chip. (a) Qubit ($Q$), resonator ($R$), flux-bias line ($\portf$), feedline input ($\porti$), and feedline output ($\porto$). (b) Zoom-in on the two Josephson junctions of the qubit. The magnetic flux threaded through the loop sets the qubit transition frequency $\fq$. (c) Zoom-in on one of the two Josephson junctions.}
\label{fig:transmon_qubit}
\end{figure}

Figure~\ref{fig:transmon_qubit} shows images at various length scales of the transmon ($Q$)~\cite{bultink2016active} that we will use in the validation. The transmon is a lumped-element nonlinear $LC$ resonator consisting of an interdigitated capacitor in parallel with a pair of Josephson junctions providing nonlinear inductance. We use the ground state (first-excited state) of this circuit as the qubit $\ket{0}$ ($\ket{1}$) state. The transition frequency $\fq$ between these states can be tuned over several gigahertz on nanosecond timescales by controlling the flux through the loop between the two Josephson junctions using the proximal flux-bias line (port $\portf$). 

Qubit measurement exploits the qubit-state dependent fundamental frequency $\fr$ of a coplanar waveguide resonator ($R$) which is capacitively coupled both to the transmon and to a feedline. A pulsed measurement (typically $300~\ns$ - $2~\us$) of transmission through the feedline (from input port $\porti$ to output port $\porto$) near the fundamental of $R$ interrogates the qubit state, projecting it to $\ket{0}$ or $\ket{1}$. Demodulation, integration, and discrimination of the transmitted signal is used to infer the measurement result.

Single-qubit gates are performed by applying calibrated microwave pulses (typically $20~\ns$) at $\fq$ to the feedline.  These pulses are commonly generated by single-sideband modulation of a carrier using an I-Q mixer and envelope functions generated by an arbitrary waveform generator. 
The envelopes and the phase of the carrier determine the rotation axis along the equator of the Bloch sphere, and the amplitude of the pulse determines the rotation angle.
Note that arbitrary single-qubit gates can be decomposed into $x$- and $y$-axis rotations albeit at the cost of longer operation sequences using some decomposition techniques, such as repeat-until-success~\cite{paetznick2014repeat}.

In circuit quantum electrodynamics, the most common two-qubit gate is the CZ gate. Such a gate can be performed between qubits coupled to a common resonator or capacitor. It is realized by applying suitably calibrated pulses of typical duration $\sim40~\ns$ to the flux-bias line. We avoid going into further detail on CZ gates here as these are not part of our validation. Please see~\cite{dicarlo2009demonstration, dicarlo2010preparation, barends2014superconducting} for details.
\section{Related Work}
\label{sec:rr}
Several quantum programming languages~\cite{omer2003structured, bettelli2003toward, green2013, wecker2014liqui, Steiger2016projectq} and compilers~\cite{svore2006layered, wecker2014liqui, javadiabhari2015scaffcc} exist in which quantum algorithms can be written and compiled into a series of instructions. These quantum compilers~\cite{svore2006layered, abhari2012scaffold, haner2016software} all generate a variant of quantum assembly language (QASM)-based instructions  that belong to the quantum instruction set. Although several quantum instruction sets have been proposed, such as a von Neumann architecture-based virtual-instruction set architecture~\cite{balensiefer2005evaluation}, quantum physical operations language (QPOL)~\cite{svore2006layered}, Hierarchical QASM with Loops (QASM-HL)~\cite{javadiabhari2015scaffcc}, Quil~\cite{smith2016practical}, and OPENQASM~\cite{ibmqasm20}, they are intermediate representations of quantum applications without considering the low-level constraints to interface with the quantum processor. They all lack an explicit control microarchitecture that implements the instructions set and allows the execution of such instructions on a real quantum processor.

Previous papers discussing quantum (micro-)~architecture can be roughly divided into three groups. The first group discusses how to physically design and fabricate a quantum processor based on a specific technology, such as trapped ions~\cite{kielpinski2002architecture, balensiefer2005evaluation, thaker2006quantum, debnath2016demonstration}, superconducting qubits~\cite{divincenzo2009fault, Brecht2016multilayer}, spin qubits~\cite{hill2015surface}, etc. The second group~\cite{oskin2002practical, metodi2005quantum, thaker2006quantum, chi2007tailoring, kreger2008microcoded, heckey2015compiler} studies how to organize qubits into multiple regions for different computational purposes to reduce the required hardware resources and communication overhead, and to maximize parallelism. The third group takes a high-level view to discuss research domains~\cite{van2013blueprint} and quantum abstraction~\cite{jones2012layered}. All of these works use the term microarchitecture differently from this paper. 

An example of control microarchitecture as viewed in this paper is~\cite{fu2016heterogeneous}, where emphasis is placed on the definition of technology-independent and technology-dependent functions in which the microcode unit plays an essential role. The microcode approach was first introduced by Wilkes~\cite{wilkes1951best} to emulate a relatively complex machine instruction as a sequence of micro-operations, called a microprogram. The microprogram can be permanently stored or cached in a control store. It enables flexible complex instruction definition using the same hardware implementation. Vassiliadis \textit{et al.}~\cite{vassiliadis2003microcode} extended the microcode method to a three-level translation from machine instructions to microinstructions and finally to micro-operations. A microinstruction decoded into one (multiple) micro-operation(s) is called vertical (horizontal). 

The microcode method is a computational model that also maps quite well onto quantum computing because: (1) there are frequently-used routines in quantum computing, such as error correction, which impact system performance significantly but can be well optimized via carefully tuning the microcode for these routines, as proposed by~\cite{kreger2008microcoded}; (2) most quantum algorithms frequently use more complex operations which cannot, at least in the foreseeable future, be directly implemented by a quantum processor. 
In this paper, we adopt the microcode approach in the proposed microarchitecture to enable flexible technology-independent instruction definition.

\section{Microarchitectural Challenges}
\label{sec:challenge}
\subsection{Motivational Example}
We use the \textit{AllXY} experiment~\cite{reed2013entanglement} as an example to illustrate the microarchitectural challenges when controlling superconducting qubits. This experiment, although simple, requires flexible control over the qubit and is sensitive to control errors such as timing inaccuracy. Hence, it can reveal some of the essential features of a microarchitecture to control a superconducting quantum processor.

\begin{figure}[hbt!]
\centering
\includegraphics[width=\linewidth]{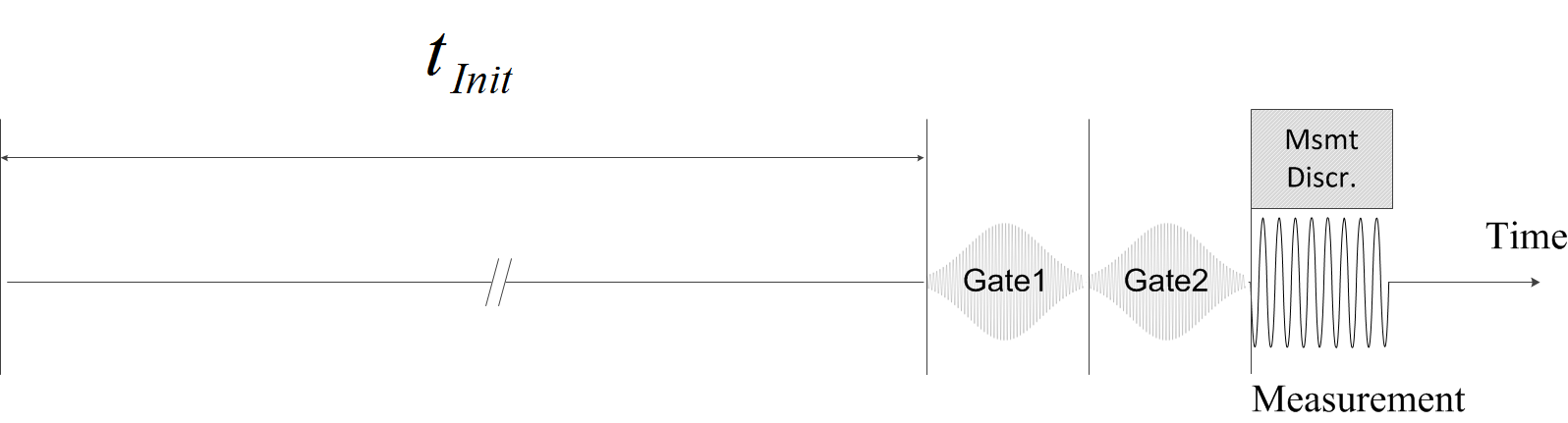}
\caption{Waveforms and timings for one round of the $AllXY$ experiment.}
\label{fig:allxy}
\end{figure}

The \textit{AllXY} experiment is a simple test of the calibration of single-qubit gates, which are realized by microwave pulses.
Different pulse errors (amplitude, frequency, etc.) produce distinct signatures that are easily recognized.
The qubit (initialized in the $\ket{0}$ state) is subjected to two back-to-back single-qubit gates and measured (Figure~\ref{fig:allxy}).
In each round, we run 21 different gate pairs: ideally, the first 5 return the qubit to $\ket{0}$, the next 12 drive it to $\frac{1}{\sqrt{2}}\left(\ket{0}+e^{in\pi/2}\ket{1}\right)$ with $n\in\{0,1,2,3\}$, and the final 4 drive it to $\ket{1}$.
By averaging the measurements results for each pair over $N$ rounds  (we take $N=25600$ in experiment), we can extract the fidelity of the qubit to the $\ket{1}$ state, and compare to the ideal staircase signature.
Algorithm~\ref{steps:allxy} shows the required procedure to perform the \textit{AllXY} experiment.

\begin{algorithm}[ht]
\SetAlFnt{\small\sf}
\KwData{gate[21][2] = \{\{$I$, $I$\}, \{$R_x(\pi)$, $R_x(\pi)$\}, \\
\quad\{$R_y(\pi)$, $R_y(\pi)$\}, \{$R_x(\pi)$, $R_y(\pi)$\}, \{$R_y(\pi)$, $R_x(\pi)$\}, \\
\quad\{$R_x(\pi/2)$, $I$\}, \{$R_y(\pi/2)$, $I$\},  \{$R_x(\pi/2)$, $R_y(\pi/2)$\}, \\
\quad\{$R_x(\pi/2)$, $R_y(\pi/2)$\}, \{$R_x(\pi/2)$, $R_y(\pi)$\}, \\
\quad\{$R_y(\pi/2)$, $R_x(\pi)$\}, \{ $R_x(\pi)$, $R_y(\pi/2)$\}, \\
\quad\{$R_y(\pi)$, $R_x(\pi/2)$\}, \{$R_x(\pi/2)$, $R_x(\pi)$\},\\
\quad\{$R_x(\pi)$, $R_x(\pi/2)$\}, \{$R_y(\pi/2)$, $R_y(\pi)$\},\\
\quad\{$R_y(\pi)$, $R_y(\pi/2)$\},\{$R_x(\pi)$, $I$\}, \{$R_y(\pi)$, $I$\}, \\
\quad\{$R_x(\pi/2)$, $R_x(\pi/2)$\}, \{$R_y(\pi/2)$, $R_y(\pi/2)$\}\};
}
 \For{($j=0$; $j < N$; $j++$)}{
    \For{($i=0$; $i < 21$; $i++$)}{
        Init the qubit; // by waiting multiple $\Tone$ ($t_{Init}$).\\
        Apply gate[i][0] on the qubit;\\
        Apply gate[i][1] on the qubit;\\
        $S_{j,i} = $ measure(qubit);
    }
 }
 $F_{\ket{1}}|_{\mathrm{meas},i} \leftarrow \sum_{j=0}^{N-1}S_{j,i}/N$;\\
 \caption{Pseudo code of the \textit{AllXY} experiment.}
 \label{steps:allxy}
\end{algorithm}

\subsection{Complex Analog Waveform Control}
\label{sec:cawc}
In classical computers, data and control signals are both binaries.
In contrast, the input and output signals of quantum processors are both complex analog signals.
The measurement outcome of qubits resides in the output analog signals from the quantum processor, 
while quantum operations on qubits (input signals) are performed by sending analog pulses that have well-defined but variable envelope, frequency, duration, timing, etc.  
For example, the $X$ gate on a transmon qubit can be implemented using a $20~\ns$ Gaussian pulse modulated to the frequency of the qubit with a particular phase.

A popular method to produce the required pulses uses arbitrary waveform generators. 
Before executing quantum algorithms, the pulses are calibrated and placed in the memory of these generators as arrays of amplitude values for each sample. 
A pulse lasting for a time $T_d$ requires the memory to store $N_s = 2\cdot T_d\cdot R_s$ samples for both in-phase (I) and quadrature (Q) components, 
where $R_s$ is the sampling rate, typically $\sim1~\mathrm{GSample/s}$. 
Each sample can consist of $\sim12$ bits, representing the vertical resolution of the amplitude.

\subsubsection{Measurement Result Discrimination}
\label{sec:mrd}
As described in Section~\ref{sec:transmon}, measurement results are contained in an analog signal $V_a(t)$. To discriminate the result for a qubit $q$, dedicated data-acquisition boards are commonly used to digitize $V_a(t)$ and perform integration and discrimination in software as follows:
\begin{align*}
S_q=\int V_a(t)W_q(t)dt,
~~\textrm{and}~~
M_q &= \begin{cases}
1 \quad \textrm{if } S_q > T_q;\\
0 \quad \textrm{otherwise.}
\end{cases}
\end{align*}
Here, $W_q(t)$ and $T_q$ are a calibrated weightfunction and threshold for $q$, respectively. $S_q$ is the integration result and $M_q$ the final binary measurement result.
The software-based method is disadvantageous because of two reasons. First, the long latency of the software-based method (hundreds of microseconds) makes real-time feedback control for superconducting qubits impossible, since latency well below the typical qubit coherence time ($<100~\us$) is required. The feedback control determines the next operations based on the result of measurements and is critical in many quantum algorithms, e.g., a specific implementation~\cite{beauregard2002circuit} of Shor's factoring algorithm~\cite{shor1994algorithms}. Second, the implied hardware resource consumption cannot scale up to a large number of qubits. A scalable measurement discrimination method with short latency constitutes a challenge.

\subsubsection{Flexible Combination of Operations}
\label{sec:flex_comb}
Quantum algorithms and even basic quantum experiments, such as \textit{AllXY}, require combining multiple quantum operations. To generate the required operation combinations, current arbitrary waveform generators first upload long waveforms combining different pulses with appropriate timing and later play them. A drawback of this method is that even a small change to the operations requires a new upload of the entire waveform which costs significant memory and upload time. To generate the 21 combinations in the \textit{AllXY} experiment, 21 different waveforms must be uploaded. With more qubits and more complex algorithms, the combination of operations can be more, which asks for more waveforms, leading to more memory consumption and larger uploading latency. Therefore, this method does not easily scale to a large number of qubits.

Furthermore, the execution of quantum programs requires more flexible feedback control, which cannot be supported by the autonomous arbitrary waveform generators as these devices cannot change a waveform to incorporate dynamically determined operations. Therefore, it is a requirement to define a flexible and scalable way to combine multiple smaller pulses, such that any sequence can be easily programmed, changed and executed when necessary.

\begin{figure*}[ht]
\centering
\includegraphics[width=\textwidth]{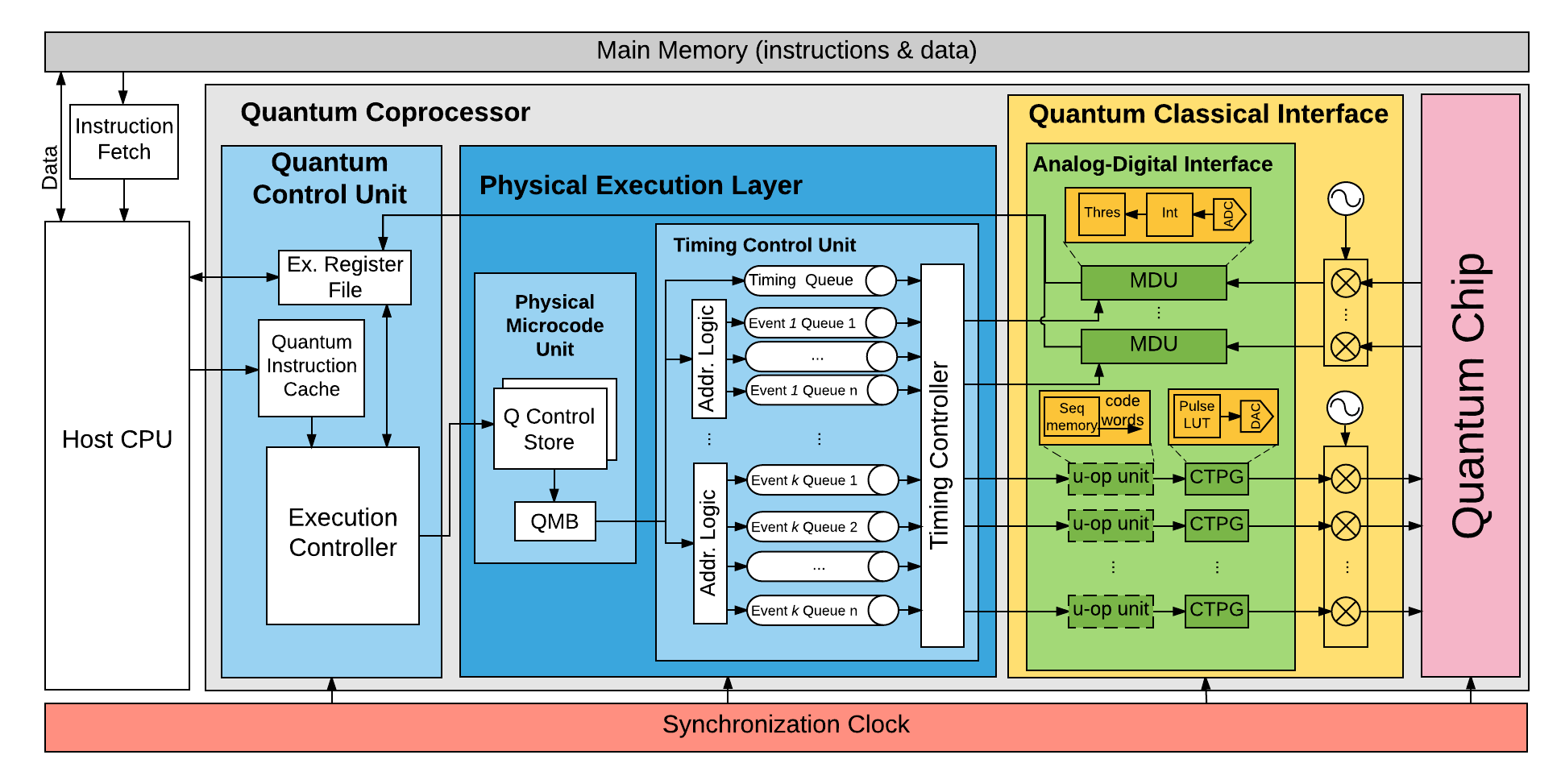}
\caption{Overview of the Quantum MicroArchitecture (QuMA).}
\label{fig:uarch_trim}
\end{figure*}

\subsubsection{Accurate Timing Control}
\label{sec:timing}
Instructions in classical processors are usually executed with non-deterministic timing on a nanosecond timescale due to (1) process switching and system calls in the software layer, (2) indefinite communication latency including memory access, (3) static and dynamical instruction reorder, (4) pipeline stall and flushing, etc. However, the non-deterministic timing typically does not matter and the program can run correctly as long as the relative order of inter-dependent instructions is preserved.

In contrast, precise timing on nanosecond timescales is critical to quantum operations. As discussed in Section~\ref{sec:transmon}, when a fixed single-sideband modulation is used, the timing of pulses must be accurate to maintain the carrier phase, which sets the rotation axis of single-qubit gates. For example, given a fixed $50~\MHz$ single-sideband modulation in the \textit{AllXY} experiment, applying the modulation envelope of an $x$ rotation $5~\ns$ later will produce a $y$ rotation instead. Besides, some quantum experiments require operations to be applied at a particular point in time. For example, the pulses implementing the two single-qubit gates and the measurement must be applied on the qubit back-to-back.
To provide the appropriate timing precision, dedicated hardware is needed where again scalability in terms of the number of qubits is an additional requirement.

Using instructions to specify the timing of operations is more promising. However, it is challenging to use non-deterministic instruction execution to generate pulses with deterministic and precise timing.

\subsection{Instruction Definition}
The instruction set architecture is the interface between hardware and software and is essential in a fully programmable classical computer. So is QISA in a programmable quantum computer.

As explained in Section~\ref{sec:rr}, existing instruction set architecture definitions for quantum computing mostly focus on the usage of the description and optimization of quantum applications without considering the low-level constraints of the interface to the quantum processor. It is challenging to design an instruction set that suffices to represent the semantics of quantum applications and to incorporate the quantum execution requirements, e.g., timing constraints.

It is a prevailing idea that quantum compilers generate technology-dependent instructions~\cite{svore2006layered, abhari2012scaffold, haner2016software}.
However, not all technology-dependent information can be determined at compile time because some information can only be generated at runtime due to hardware limitations. 
An example is the presence of defects on a quantum processor affecting the layout of qubits used in the algorithm. In addition, the following observations hold: (1) quantum technology is rapidly evolving, and more optimized ways of implementing the quantum gates are continuously explored and proposed; a way to easily introduce those changes, without impacting the rest of the architecture, is important.
(2) depending on the qubit technology, the kind, number and sequence of the pulses can vary. Hence, it forms another challenge to microarchitecturally support a set of quantum instructions which is as independent as possible of a particular technology and its current state of the art.
\section{Quantum Microarchitecture}
\label{sec:uarch}
In this section, we describe the Quantum MicroArchitecture (QuMA) as shown in Figure~\ref{fig:uarch_trim}. QuMA is a heterogeneous architecture which includes a classical CPU as a host and a quantum coprocessor as an accelerator. 

As proposed in~\cite{fu2016heterogeneous}, the input of QuMA is a binary file generated by a compiler infrastructure where classical code and quantum code are combined. The classical code is produced by a conventional compiler such as GCC and executed by the classical host CPU. Quantum code is generated by a quantum compiler and executed by the quantum coprocessor.

As shown in Figure~\ref{fig:uarch_trim}, the host CPU fetches quantum code from the memory and forwards it to the quantum coprocessor. In the quantum coprocessor, executed instructions in general flow through modules from left to right. The execution controller performs register update, program flow control and streams quantum instructions to the physical execution layer. 
The physical microcode unit translates quantum instructions into microinstructions using the Q control store. These are further decomposed into micro-operations by the quantum microinstruction buffer (QMB). The timing of each micro-operation is also determined by the physical microcode unit. 
Based on the output of quantum microinstruction buffer, the timing control unit triggers micro-operations at a deterministic timing.
The analog-digital interface converts digitally represented micro-operations into corresponding analog pulses with precise timing that perform quantum operations on qubits, as well as analog signals containing measurement information of qubits into binary signals. Required modulation and demodulation with radio-frequency carrier waves are also carried out in the quantum-classical interface.

In order to address the challenges described in the previous section, three schemes are introduced in QuMA. (i) The codeword-based event control scheme is implemented by the codeword-triggered pulse generation unit (CTPG), which produces analog input to the quantum processor based on the received codeword triggers, and the measurement discrimination unit (MDU) converting the analog output from the quantum processor into binary results.
(ii) The queue-based event timing control scheme is implemented by the timing control unit, which issues event triggers with precise timing to the measurement discrimination unit and the micro-operation unit (u-op unit). (iii)  A multilevel instruction decoding scheme, which successively decodes a quantum instruction into microinstructions at the Q Control Store, micro-operations at the quantum microinstruction buffer, and finally codeword triggers at the micro-operation unit. The complex analog waveform control challenge is addressed by (i) and (ii) whereas the instruction definition is addressed by (iii).

\subsection{Codeword-Based Event Control}
\label{sec:adi}
The analog-digital interface (Figure~\ref{fig:uarch_trim}) is at the boundary of analog signals and digital signals in QuMA, which is technology-dependent. As shown in Figure~\ref{fig:uarch_trim}, from left to right , the micro-operation unit and the codeword-triggered pulse generation unit translate codeword triggers into pulses representing quantum operations on the qubits with a fixed latency. From right to left, analog measurement waveforms from the quantum processor are discriminated into binary results by the measurement discrimination unit.  In this way, the analog-digital interface abstracts the complex analog waveform generation and puts forward the responsibility of codeword control with precise timing to the upper digital layers. Therefore, it enables controlling analog pulse generation using instructions. Fast and flexible feedback control is also possible in principle because the codeword-triggered pulse generation scheme does not require the waveform to be uploaded at runtime and codeword triggers with precise timing can be efficiently generated dynamically.

\begin{figure*}[t]
\centering
\includegraphics[width=\textwidth]{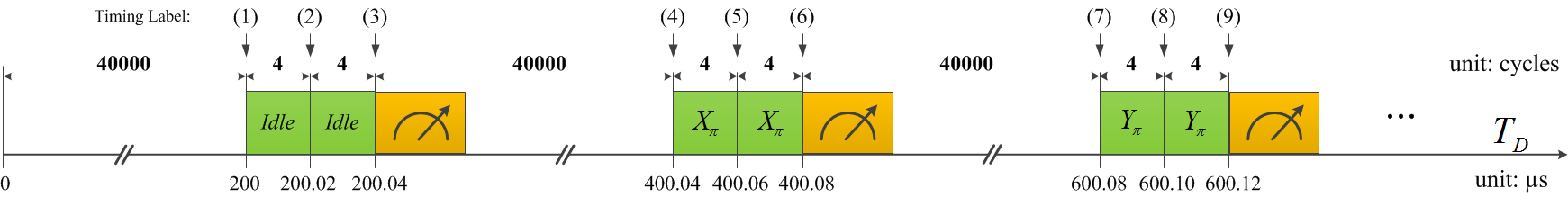}
\caption{Operations of the \textit{AllXY} experiment in the timeline. Measurement pulse generation and measurement result discrimination overlap in time and are shown using the same meter box.}
\label{fig:allxy_timeline}
\end{figure*}

\subsubsection{Codeword-Triggered Pulse Generation}
\label{sec:cptg}
From experiments, we observe that the pulses for a fixed and small set of quantum operations can be well defined and used after calibration. They are also called primitive operations because they are sufficient for many quantum computing experiments. Based on this, we introduce the codeword-triggered pulse generation scheme in QuMA to generate pulses corresponding to primitive operations. In codeword-triggered pulse generation, well-defined primitive pulses instead of entire waveforms are uploaded to the memory. The memory is organized as a lookup table and each entry in the lookup table, indexed by means of a codeword, contains the sample amplitudes corresponding to a single pulse. The codeword-triggered pulse generation unit converts a digitally stored pulse into an analog one only when it receives a codeword trigger. An example of the lookup table content for single-qubit operations is shown in Table~\ref{tab:lut}.

\begin{table}[ht]
\caption{An example of the lookup table content of a codeword-triggered pulse generation unit for single-qubit gates.}
    \centering
    \small
\begin{tabular}{|c|c|c|c|c|}
    \hline
    Codeword & 0 & 1 & 2 & 3  \\ \hline
    Pulse & $I$ & $R_x(\pi)$ & $R_x(\frac{\pi}{2})$ & $R_x(-\frac{\pi}{2})$\\ \hline
    Codeword & 4 & 5 & 6 & $\cdots$ \\ \hline
    Pulse &  $R_y(\pi)$ & $R_y(\frac{\pi}{2})$ & $R_y(-\frac{\pi}{2})$ & $\cdots$\\ \hline
\end{tabular}
\label{tab:lut}
\end{table}

The codeword-triggered pulse generation scheme has a modest memory requirement since it only needs to store a small number of pulses for the well-defined primitive operations. In the \textit{AllXY} experiment, only the pulses for 7 operations need to be stored, which only consumes the memory for $7\times2\times20~\ns\times R_s$ samples (in total 420 Bytes), instead of 21 waveforms each containing two operations, that are $21\times2\times2\times20~\ns\times R_s$ samples (in total 2520 Bytes).
When more complex combination of operations is required, the memory consumption will remain the same and the memory saving will be more significant. The small memory footprint provides a scalable path for controlling a larger number of qubits.

The delay between the codeword trigger and the pulse generation is required to be fixed and short in the codeword-triggered pulse generation unit. The fixed delay ensures that the flexible combination of the pulses with precise timing can be achieved by flexibly generating the corresponding codeword triggers at precise timing. In the \textit{AllXY} experiment, by issuing the codeword triggers for the two gates with an interval of $20~\ns$, the pulses for the two gates can be played out exactly back to back. 

\subsubsection{Measurement Discrimination}
\label{sec:msmt}
Recent experiments have demonstrated measurement discrimination using a customized FPGA~\cite{bultink2016active}, achieving a short latency $<1~\us$ which enables real-time feedback control. This method also costs modest hardware exhibiting better scalability. Adopting this idea, we introduce hardware-based measurement discrimination units in the analog-digital interface. The measurement discrimination unit translates the analog signal containing measurement information of a single qubit into a binary measurement result. Once the measurement discrimination unit for qubit $q$ receives a codeword trigger, it starts the measurement discrimination process and generates a binary result $R_q$. $R_q$ can be subsequently forwarded to the quantum control unit for feedback control or reading back.

Recent experiments have also demonstrated combining the measurement result of multiple qubits into one analog signal~\cite{riste2015detecting, ryan2017hardware}. This can reduce the number of required measurement discrimination units and exhibits better scalability.

\subsection{Queue-Based Event Timing Control}
\label{sec:quetic}
The timing control unit divides the microarchitecture into two timing domains: the non-deterministic timing domain and the deterministic timing domain, which are on the left and right side of the timing control unit in Figure~\ref{fig:uarch_trim}, respectively. In the non-deterministic timing domain, the quantum control unit and physical execution layer execute instructions and feed quantum operations to the queues in an as-fast-as-possible fashion. In the deterministic timing domain, quantum operations in the queue are emitted to the analog-digital interface with deterministic and precise timing. To this end, queue-based event timing control is introduced.

To illustrate the working principle of queue-based event timing control, the operations of the \textit{AllXY} experiment with corresponding timing are shown in Figure~\ref{fig:allxy_timeline}. The horizontal axis labels mark the time points in microseconds when a corresponding operation takes place. Each time point is assigned a timing label, which is the number in brackets on the top. The bold numbers above the double-arrow lines indicate intervals between two time points in cycles. Here and throughout the rest of the paper, a cycle time of $5~\ns$ is used.

The timing control unit implements queue-based event timing control in QuMA. It consists of a timing queue, multiple event queues, and a timing controller. 
The timing queue buffers the time points with corresponding timing labels. 
The location of the time points can be designated in the timeline, e.g., by specifying the intervals between consecutive time points as shown in Figure~\ref{fig:allxy_timeline} and the first column of Table~\ref{tab:t1_zero}.
Each event queue buffers a sequence of events with a time point at which the event is expected to take place. The time point is indicated by the aforementioned timing label. An event can be a quantum gate, measurement, or any other operation.  
The timing controller maintains the clock of the deterministic timing domain ($T_D$), which can be started by an instruction or another source, e.g., an external trigger. When $T_D$ reaches the assigned time point, the timing controller signals the queues to fire the events matching that time point and emits them to the analog-digital interface.

\begin{table}[!t]
\begin{minipage}[t]{\linewidth}
    \small
    \centering
    \captionof{table}{Queue state of the \textit{AllXY} experiment when $T_D=0$.}
    \label{tab:t1_zero}
    \begin{tabular}{|c|c|c|c|}
    \hline
       Timing Queue &   Pulse Queue             & MPG Queue                  & MD Queue                  \\ \hline \hline
       $\vdots$     & \multirow{3}{*}{$\vdots$} & \multirow{4}{*}{$\vdots$} & \multirow{4}{*}{$\vdots$} \\ \cline{1-1}
       (4, 6)       &                           &                           &                           \\ \cline{1-1}
       (4, 5)       &                           &                           &                           \\ \cline{1-2}
       (40000, 4)   & ($X_\pi$, 5)              &                           &                           \\ \cline{1-2}
       (4 , 3)      & ($X_\pi$, 4)              &                           &                           \\ \hline
       (4, 2)       & ($I$, 2)        & (6)                       & ($r7$, 6)                \\ \hline
       (40000, 1)   & ($I$, 1)        & (3)                       & ($r7$, 3)                \\ \hline
    \end{tabular}
    \vspace{10pt}
    \captionof{table}{Queue state of the \textit{AllXY} experiment when $T_D=40000$. }
    \begin{tabular}{|c|c|c|c|}
    \hline
       Timing Queue &   Pulse Queue             & MPG Queue                  & MD Queue                  \\ \hline \hline
       $\vdots$     & \multirow{3}{*}{$\vdots$} & \multirow{4}{*}{$\vdots$} & \multirow{4}{*}{$\vdots$} \\ \cline{1-1}
       (4, 6)       &                           &                           &                           \\ \cline{1-1}
       (4, 5)       &                           &                           &                           \\ \cline{1-2}
       (40000, 4)   & ($X_\pi$, 5)              &                           &                           \\ \hline
       (4, 3)       & ($X_\pi$, 4)              & (6)                       & ($r7$, 6)                \\ \hline
       (4, 2)       & ($I$, 2)        & (3)                       & ($r7$, 3)                \\ \hline
    \end{tabular}
    \label{tab:t1_one}
    \vspace{10pt}
    \captionof{table}{Queue state of the \textit{AllXY} experiment when $T_D=40008$. }
    \begin{tabular}{|c|c|c|c|}
    \hline
       Timing Queue &   Pulse Queue             & MPG Queue                  & MD Queue                  \\ \hline \hline
       $\vdots$     & \multirow{2}{*}{$\vdots$} & \multirow{3}{*}{$\vdots$} & \multirow{3}{*}{$\vdots$} \\ \cline{1-1}
       (4, 6)       &                           &                           &                           \\ \cline{1-2}
       (4, 5)       & ($X_\pi$, 5)              &                           &                           \\ \hline
       (40000, 4)   & ($X_\pi$, 4)              & (6)                       & ($r7$, 6)                \\ \hline
    \end{tabular}
    \label{tab:t1_two}
\end{minipage}
\end{table}

In order to better illustrate how queue-based event timing control works, we use the \textit{AllXY} experiment. Three event queues are used in this experiment (see Table~[2-4]): the Pulse Queue for single-qubit operations, the MPG Queue for measurement pulse generation, and the MD Queue for measurement discrimination. 
Besides the timing label for each event, the pulse queue contains the single-qubit operations, e.g., the $I$ or $X_\pi$ operation, to be triggered, and the MD queue contains the destination register, e.g., $r7$, to write back the measurement result. After executing a couple of instructions in the program and before $T_D$ is started, the state of the queues is as shown in Table~\ref{tab:t1_zero}. The bottom of the table corresponds to the front of the queues.
After $T_D$ is started, a counter in the timing controller starts counting. When the counter reaches the first interval value in the timing queue, i.e., 40000, the corresponding timing label, i.e., 1, is broadcast to all event queues. At the same time, the counter resets and restarts. 
Since the pulse queue contains that same label, 1, at the front of the queue, the operation $I$ is fired to the analog-digital interface. The queue state then turns into Table~\ref{tab:t1_one}. The second $I$ operation is issued in the same way when the counter reaches the next interval value, 4.
After the counter reaches the third interval value, 4, the timing label 3 is broadcast and the MG Queue triggers the measurement pulse generation and the MD queue triggers a measurement discrimination process of which both associated timing labels are 3. The queue state then turns into Table~\ref{tab:t1_two}. The rest can be done in the same manner.

\subsection{Multilevel Instruction Decoding}
\label{sec:multidecoding}
Combining the codeword-based event control scheme and queue-based event timing control enables other stages in QuMA to focus on flexibly decoding the quantum instructions and filling the queues as fast as possible without worrying about complex analog waveform control with rigid timing constraints. In this subsection, we first give an overview of the instruction definition and then discuss the multilevel decoding scheme for the quantum instructions.

\subsubsection{Instruction Definition}
The quantum code is written with instructions in the Quantum Instruction Set (QIS). An example of QIS instructions is shown in Table~\ref{tab:instr_level}. QIS contains auxiliary classical instructions and quantum instructions. Auxiliary classical instructions are used for basic arithmetic and logic operations and program flow control. Quantum instructions describe which and when quantum operations will be applied on qubits. By including auxiliary classical instructions, QIS can support feedback control based on measurement results and a hierarchical description of quantum algorithms which can significantly reduce the program code size~\cite{kudrow2013quantum}. 

\subsubsection{Instruction Decoding}
\label{sec:mld}
To support a technology-independent quantum instruction set definition, we adopt a multilevel instruction decoding approach in which quantum instructions, especially that for quantum gates, are successively decoded into quantum microinstructions, micro-operations and finally codeword triggers to control codeword-triggered pulse generation  to generate pulses. For example, Table~\ref{tab:instr_level} shows four decoding steps for the instructions of the \textit{AllXY} experiment. From the QIS on, time is calculated in cycles. Due to the simplicity of the \textit{AllXY} experiment and for the sake of code efficiency, the inner loop as shown in Algorithm~\ref{steps:allxy} is unrolled. The execution of quantum instructions starts from the execution controller.

\begin{table}[hbt]
\centering
\small
\caption{The format of QIS instructions, quantum microinstructions, micro-operations and codeword triggers. Taking the \textit{AllXY} experiment as an example.}
\begin{tabular}{|l|l|}
\hline
\multicolumn{1}{|c|}{\textbf{QIS}}  & \multicolumn{1}{c|}{\textbf{QuMIS}} \\ \hline
\begin{tabular}[t]{@{}l@{}}
\gray{\# Input to the execution controller}\\
\quad \quad mov r1, 0\\
\quad \quad mov r2, 25600\\
\quad \quad mov r3, ResultMemAddr\\
\quad \quad mov r15, 40000\\
Outer\_Loop: \\
\quad \quad QNopReg r15\\
\quad \quad Apply I, q0\\
\quad \quad Apply I, q0\\
\quad \quad Measure q0, r7 \\
\quad \quad Load r9, r3[0]\\
\quad \quad Add r9, r9, r7\\
\quad \quad Store r9, r3[0]\\
\\
\quad \quad QNopReg r15\\
\quad \quad Apply X180, q0\\
\quad \quad Apply X180, q0\\
\quad \quad Measure q0, r7 \\
\quad \quad Load r9, r3[1]\\
\quad \quad Add r9, r9, r7\\
\quad \quad Store r9, r3[1]\\
\quad \quad ...\\
\quad \quad add r1, r1, 1\\
\quad \quad bne r1, r2, Outer\_Loop\\\end{tabular} & 
     
\begin{tabular}[t]{@{}l@{}}
\gray{\# Input to the QMB}\\
\gray{\# round 0:}\\
\quad \quad Wait 40000\\
\quad \quad Pulse \{q0\}, I\\
\quad \quad Wait 4\\
\quad \quad Pulse \{q0\}, I\\
\quad \quad Wait 4\\
\quad \quad MPG \{q0\}, 300\\
\quad \quad MD \{q0\}, r7\\
\gray{\# round 1:}\\
\quad \quad Wait 40000\\
\quad \quad Pulse \{q0\}, X180\\
\quad \quad Wait 4\\
\quad \quad Pulse \{q0\}, X180\\
\quad \quad Wait 4\\
\quad \quad MPG \{q0\}, 300\\
\quad \quad MD \{q0\}, r7\\
\quad$\ldots$\\
\end{tabular} \\ \hline\hline

\multicolumn{1}{|c|}{\textbf{Micro-operations}}  & \multicolumn{1}{c|}{\textbf{Codeword Triggers}} \\ \hline

\begin{tabular}[t]{@{}l@{}}
\gray{\# Input to the u-op units}\\
\\
$T_D=40000$:\\
~~  $I$ sent to u-op unit0\\
$T_D=40004$:\\
~~  $I$ sent to u-op unit0\\
$T_D=40008$:\\ 
\gray{~~\# MPG\&MD bypass this stage}\\
$T_D=80008$: \\
~~  $X_\pi$ sent to u-op unit0\\
$T_D=80012$: \\
~~  $X_\pi$ sent to u-op unit0\\
$T_D=80016$: \\
\gray{~~\# MPG\&MD bypass this stage}\\
\quad$\ldots$\\

\end{tabular}              & 

\begin{tabular}[t]{@{}l@{}}
\gray{\# Input to the MDU or CPTG}\\
\gray{\# $\Delta$ is the delay of the u-op unit}\\

$T_D=40000+\Delta$:\\
~~   $CW~0$ sent to CTPG0\\
$T_D=40004+\Delta$:\\
~~   $CW~0$ sent to CTPG0\\
$T_D=40008$:\\
~~   $CW~7$ sent to CTPG5 \gray{\# Msmt}\\
~~   \textit{MD(r7)} sent to MDU0\\
$T_D=80008+\Delta$:\\
~~   $CW~1$ sent to CTPG0\\
$T_D=80012+\Delta$:\\
~~   $CW~1$ sent to CTPG0\\
$T_D=80016$:\\
~~   $CW~7$ sent to CTPG5 \gray{\# Msmt}\\
~~   \textit{MD(r7)} sent to MDU0\\
~~   $\ldots$\\
\end{tabular}\\ \hline
\end{tabular}
\label{tab:instr_level}
\end{table}

\paragraph{Execution Controller}
This unit executes the auxiliary classical instructions in the QIS and streams quantum instructions to the physical microcode unit.
By executing the auxiliary classical instructions in the execution controller, the same quantum instruction can be issued to the physical microcode unit multiple times and each time with expected parameters computed at runtime. For example, the \textit{QNopReg r15} instruction in the QIS is used to specify the initialization time. Each of the 21 \textit{QNopReg r15} instructions will be issued once per round. Every time it is issued, it reads a waiting time from the register r15, which results in a \textit{Wait 40000} instruction. If the register value is updated using auxiliary classical instructions, the waiting time specified in the \textit{Wait} instruction can be calculated at runtime. In this way, it enables a compact and flexible description of quantum algorithms.

\paragraph{Physical Microcode Unit}
Quantum instructions are translated into a sequence of microinstructions in the physical microcode unit based on the microprograms uploaded into the Q control store. The timing for each quantum operation is also determined at this stage.
For now and as shown in Table~\ref{tab:qumis}, the microinstruction set, QuMIS, consists of the following instructions: i) the \textbf{Wait} instruction used to specify the interval between consecutive time points, ii) the \textbf{Pulse} instruction used to apply quantum gates on qubits; iii) the \textbf{MPG} instruction used to generate the measurement pulse; iv) the \textbf{MD} instruction used to trigger the measurement discrimination process.

In the quantum microinstruction buffer (QMB), quantum microinstructions for quantum gates are decomposed into separate micro-operations with timing labels and push them into the queues in the timing control unit as shown in Table~\ref{tab:t1_zero}. Due to the simplicity of measurements in terms of instruction control, quantum microinstructions for measurement pulse generation or measurement discrimination can be directly translated into codeword triggers to control the codeword-triggered pulse generation unit or the measurement discrimination unit bypassing the micro-operation unit. The timing control unit then emits the micro-operations at the expected timing. The \textbf{Pulse} and \textbf{MPG} instructions are both horizontal instructions, which can trigger the operation on multiple qubits at the same time.

\begin{table}[t]
\footnotesize
\centering
\caption{QuMIS instructions.}
\label{tab:qumis}
\begin{tabular}{|c|l|}
\hline
\textbf{\begin{tabular}[c]{@{}c@{}}Assembly Format\end{tabular}} & \multicolumn{1}{c|}{\textbf{Description}} \\ \hline
Wait \textit{Interval} & \begin{tabular}[c]{@{}l@{}}Wait for the number of cycles indicated\\by the immediate value \textit{Interval}.\end{tabular} \\ \hline
\begin{tabular}[c]{@{}l@{}} Pulse $(QAddr_0, uOp_0)\mathbf{\left[\right.}$, \\~~$(QAddr_1, uOp_1), \ldots \mathbf{\left.\right]}$  \end{tabular} & \begin{tabular}[c]{@{}l@{}}Apply the micro-operation $uOp_i$ on each\\ of the qubit(s) specified by the address\\$QAddr_i$.\end{tabular} \\ \hline
MPG \textit{QAddr}, \textit{D} & \begin{tabular}[c]{@{}l@{}} Generate the measurement pulse for\\ the qubits specified by the address \textit{QAddr}.\\ \textit{D} indicates the duration of the\\ measurement pulse in number of cycles. \end{tabular} \\ \hline
MD \textit{QAddr}, \textit{\$rd}& \begin{tabular}[c]{@{}l@{}} Discriminate the measurement results of\\ the qubits specified by \textit{QAddr} and store\\ the result into register \textit{\$rd}.\end{tabular} \\ \hline
\end{tabular}
\end{table}

Let us illustrate these concepts using the CNOT gate. A CNOT gate with a control qubit $c$ and a target qubit $t$ can be decomposed in the following way~\cite{nielsen2010quantum}:
\begin{align*}
    \textrm{CNOT}_{c,t}=R_y(\pi/2)_t\cdot CZ \cdot R_y(-\pi/2)_t.
\end{align*}
Adopting the microcoded approach for the instruction \textit{CNOT qt, qc} applying on superconducting qubits results in Algorithm~2. 
\begin{algorithm}[h!]
\lstset{language=QuMIS}
\begin{lstlisting}
Pulse   {qt},     Ym90
Wait    4
Pulse   {qt, qc}, CZ
Wait    8
Pulse   {qt},     Y90
Wait    4
\end{lstlisting}
\label{alg:cnotgate}
\caption{Microprogram for the physical \textit{CNOT q1, q2}.}
\end{algorithm}

By utilizing horizontal microcode, one quantum instruction can be translated into multiple microinstructions and one microinstruction into multiple micro-operations. This allows flexible emulation of complex, technology-independent instructions using technology-dependent primitives. 

\paragraph{Micro-Operation Unit}
At the micro-operation unit, each micro-operation is translated into a sequence of codeword triggers with predefined latency, which further makes associated codeword-triggered pulse generation units generate primitive operation pulses.
For each predefined micro-operation $uOp_i$, the micro-operation unit stores a sequence $\mathrm{Seq}_i$ comprising of codewords and timing. $\mathrm{Seq}_i$ has the following format:
\begin{equation*}
\mathrm{Seq}_i: ~~([0,~cw_0]; ~~[\Delta t_{1},~cw_1]; ~~[\Delta t_{2},~cw_2]; ~~\ldots),
\end{equation*}
where $\Delta t_{j}$ represents the interval between codeword triggers $cw_{j-1}$ and $cw_j$. 
Once the micro-operation $uOp_i$ is triggered, the micro-operation unit starts to output codeword $cw_j$ after waiting for $\Delta t_{j}$ cycles sequentially as defined in the sequence $\mathrm{Seq}_i$. Since the timing controller fires the micro-operation at precise timing, the codeword triggers are also generated at precise timing.

For example, a $Z$ gate can be decomposed into a $Y$ gate followed by an $X$ gate since $Z = X \cdot Y$ (up to an irrelevant global phase). The micro-operation unit can perform the translation for superconducting qubits using the following sequence given the lookup table content as listed in Table~\ref{tab:lut}:
\begin{equation*}
\mathrm{Seq}_Z: ~~([0,~1]; ~~[4,~4]).
\end{equation*}

The micro-operation unit allows the emulation of commonly-used quantum operations which are not directly implementable using primitive operations. Moreover, it reduces the communication between the timing control unit and the analog-digital interface. This is especially helpful when the timing control unit and the analog-digital interface are implemented in different electronic devices for performance and scalability.

\section{Evaluation}
\label{sec:scale}
To evaluate QuMA, we make a comparison between QuMA and the architecture of the Raytheon BBN APS2 system, which is a commercial device that has been recently demonstrated~\cite{ryan2017hardware, bbnaps2}
for superconducting qubits. Then we discuss the scalability limitation of QuMA.

The APS2 system has a distributed architecture consisting of nine individual APS2 modules and a trigger distribution module (TDM) that can fully control up to eight qubits. A quantum application is translated into multiple binary executables running in parallel on each of the APS2 modules. A binary is composed of separated program flow control instructions and output instructions.
Instead of instructions with explicit quantum semantics, low-level output instructions are used, such as waveform with a physical memory address. 
Idle waveforms are used to implement precise timing between operations, and the TDM distributes trigger signals to perform parallelism/synchronization of multiple outputs via an interconnect network.
The main disadvantage are that no output instructions can be processed when synchronization is required, and the interconnect network is cumbersome and fragile when scaling up to tens of qubits where multiple APS2 systems are required~\cite{ryan2017hardware}.

In contrast, QuMA employs a centralized architecture, in which: (i) only one binary executable is required for controlling multiple qubits, (ii) quantum semantics and timing of operations are explicitly defined at the instruction level, (iii) parallelism/synchronization of outputs is achieved by triggering events at specific timing points, which is neither dependent on another module nor limited by the interconnect network.
These three points contribute to a relatively simple compilation model for QuMA. 
As explained in Section~\ref{sec:quetic}, QuMA decouples the timing of executing instructions and performing output. So it can maintain fully deterministic timing of the output and maximally process instructions during waiting. 
Since data is gathered in a single place (the register file), it is natural to extend QuMA to a heterogeneous computing platform by adding extra data exchange instructions to interact with the host CPU and the main memory.

\setcounter{figure}{6}
\begin{figure*}[tb]
\centering
\includegraphics[width=\linewidth]{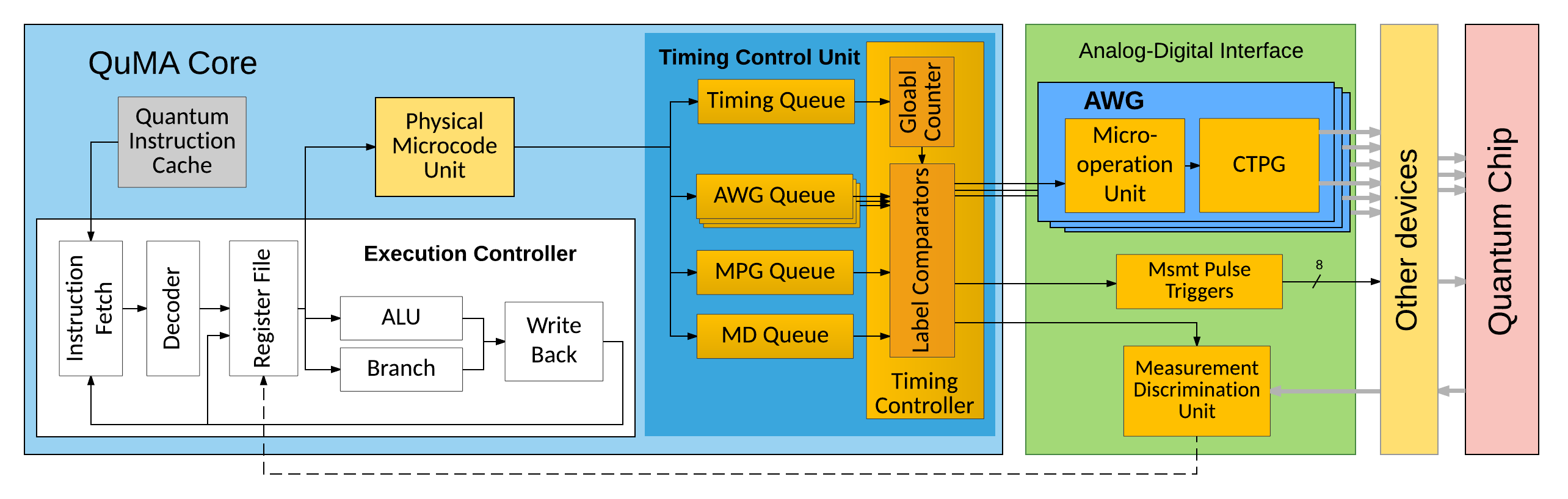}
\caption{Schematic of the implemented QuMA. The thick gray lines are analog signals while the dark thin lines are digital signals. Dashed lines indicate functionality to be added in the future.}
\label{fig:quma_core}
\end{figure*} 

Regarding scalability, QuMA is not limited by the analog-digital interface and the timing control unit, as their size scales linearly to the number of qubits and can be implemented in a distributed way. 
However, the limited time for executing instructions in quantum computers may form a challenge in QuMA when more qubits ask for a higher operation output rate while only a single instruction stream is used.
A Very-Long-Instruction-Word (VLIW) architecture~\cite{Hennessy2011computer} can be adopted to provide much larger instruction issue rate. In addition, by optimizing the microcode unit and the micro-operation unit, it is possible to use less quantum instructions to describe more quantum operations, which can relax the instruction issue rate requirement.

\setcounter{figure}{5}
\begin{figure}[tb]
\centering
\includegraphics[width=\linewidth]{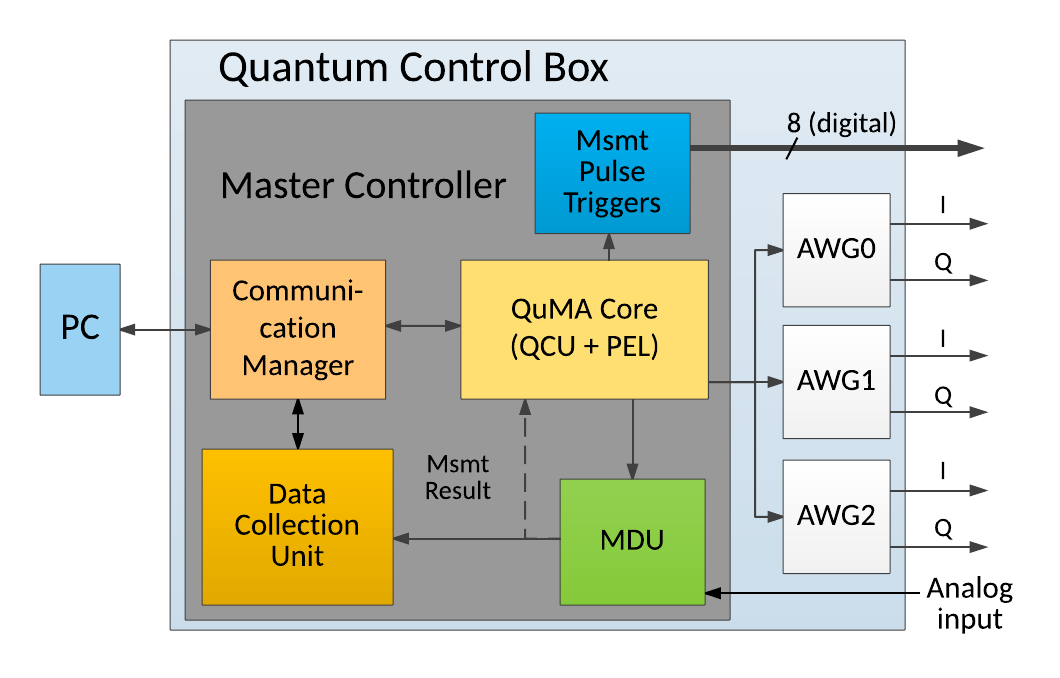}
\caption{Schematic of the CBox firmware architecture. The QuMA core is implemented in the Master Controller. Dashed lines indicate functionality to be added in the future.}
\label{fig:cbox3}
\end{figure}
\setcounter{figure}{7}
\section{Implementation}
\label{sec:implementation}
In this section, we discuss the quantum control box, where the aforementioned mechanisms have been implemented. 

\subsection{Quantum Control Box}
\label{subsec:cbox}
The quantum control box, as shown schematically in Figure~\ref{fig:cbox3}, consists of four FPGA boards. One board implements the Master Controller and the other three boards implement a two-channel arbitrary waveform generator (AWG) each. 

The master controller is implemented using an Arrow BeMicro CV A9 board holding an Altera Cyclone V 5CEFA9 FPGA chip. 
It connects to two 8-bit resolution analog-to-digital converters (ADC) that digitize analog measurement signals from the quantum chip. 
The master controller has eight digital outputs used for triggering measurement pulse generation and triggers the pulse generation of each AWG via a pair of Low-Voltage-Differential-Signaling wires.

Inside the MC, the QuMA core implements the quantum control unit and the physical execution layer of QuMA. 
The digital output unit converts the measurement operation tuple ($QAddr$, $D$) received from the QuMA core into `1' state with a duration of $D$ cycles for the eight digital outputs masked by $QAddr$.
The measurement discrimination unit (MDU) can discriminate the measurement result of a single qubit. The data collection unit can collect $K$ consecutive integration results of a single qubit for $N$ rounds, calculate and store the average of $K$ integration results across the $N$ rounds:
\begin{align*}
\bar{S_i} = \left(\sum_{j=0}^{N-1}S_{i,j}\right)/N ~,~~ i\in\{0, 1, \cdots, K-1\}.
\end{align*}
After the data collection process is done, the PC can retrieve the averaging integration results $\{\bar{S_i}\}$.

Each AWG is implemented using a Terasic DE0-Nano board holding an Altera Cyclone IV EP4CE22F FPGA chip and uses two 14-bit resolution digital-to-analog converters (DAC) to generate the in-phase and quadrature components of qubit control pulses. Each AWG includes a micro-operation unit and a codeword-triggered pulse generation unit. The implemented codeword-triggered pulse generation unit has a fixed delay of $80~\ns$ from the codeword trigger to the output pulse.

All FPGAs, ADCs, and DACs are clocked at 200~MHz, except for communication and data collection, which run at 50~MHz. The MC communicates with the PC via USB. The MC communicates to the AWGs, e.g., uploading the lookup table content of the codeword-triggered pulse generation unit.

\subsection{QuMA Implementation}
The QuMA implementation in the  control box in shown in Figure~\ref{fig:quma_core}. In view of the running physics experiments, it slightly differs from the microarchitecture presented in Section~\ref{sec:uarch}. We have partially implemented the system including the quantum instruction cache, the execution controller, part of the physical microcode unit, the timing control unit and the quantum classical interface. The rest is planned for future release.
Due to the absence of a fully functioning physical microcode unit, the high-level quantum instructions of the QIS are not implemented yet. A combination of the auxiliary classical instructions in the QIS and QuMIS (see Table~\ref{tab:qumis}) is loaded into the quantum instruction cache.

We have designed a quantum programming language OpenQL based on C++ with a compiler that can translate the OpenQL description into the auxiliary classical instructions and QuMIS instructions.

The execution controller incorporates a classical pipeline to execute auxiliary classical instructions. The register file in this pipeline contains runtime information related to quantum program execution. QuMIS instructions are dispatched to the physical microcode unit after reading register values. The physical microcode unit can determine the timing of QuMIS instructions and decompose QuMIS instructions into micro-operations. A full implementation of the physical microcode unit is still under development. The timing control unit implements the queue-based event timing control scheme (as described in Section~\ref{sec:quetic}). 
The measurement pulse triggers pulse modulated microwave carrier generators in the other devices block to produce the measurement pulse for qubits.
\section{Experimental Results}
\label{sec:exp}
\begin{figure}[t!]
\centering
\includegraphics[width=\linewidth]{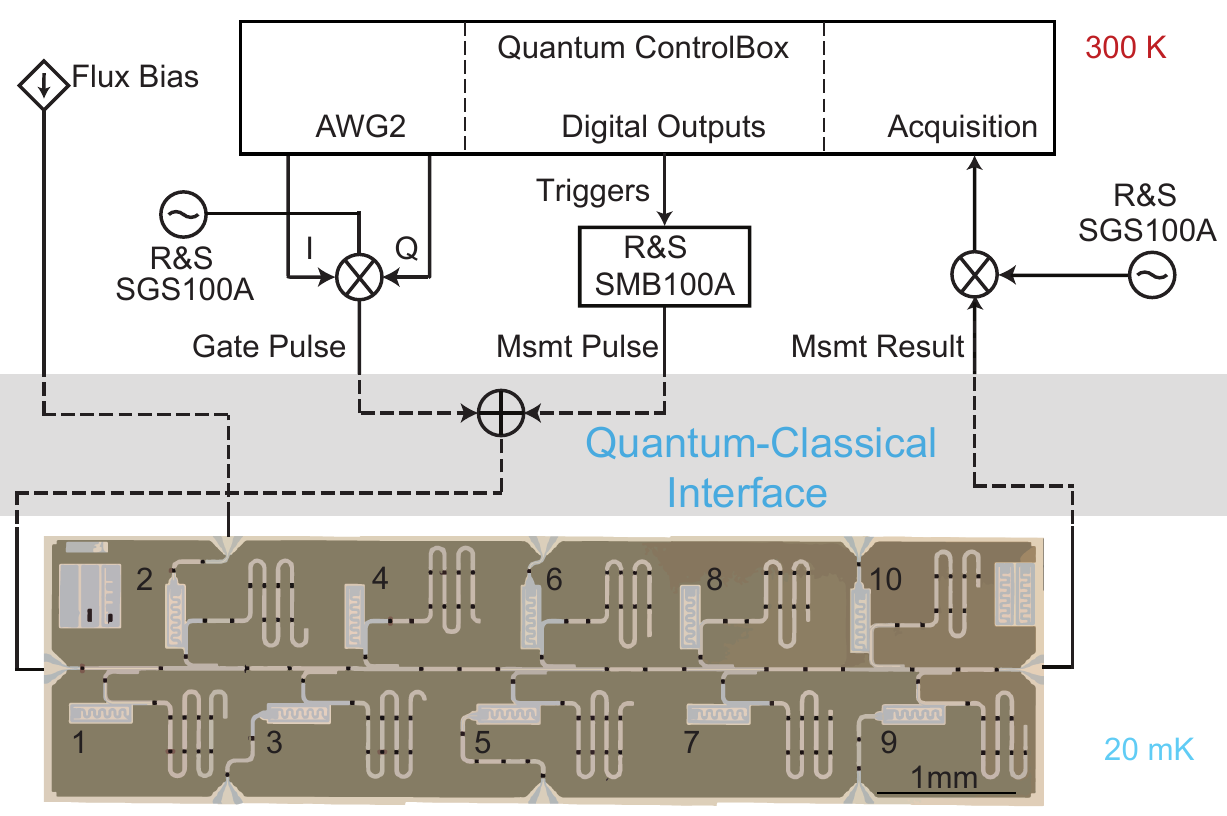}
\caption{Experimental setup used for validation of the microarchitecture. }
\label{fig:setup}
\end{figure}

\begin{algorithm}[h!]
\lstset{language=QuMIS}
\begin{lstlisting}
mov  r15,  40000     # 200 us
mov  r1,   0         # loop counter
mov  r2,   25600     # number of averages

Outer_Loop:
  QNopReg r15        # Identity, Identity
  Pulse   {q2}, I
  Wait    4
  Pulse   {q2}, I
  Wait    4
  MPG     {q2}, 300
  MD      {q2}
(repeat the previous 7 instructions once again)

  QNopReg r15        # X180, X180
  Pulse   {q2}, X180
  Wait    4
  Pulse   {q2}, X180
  Wait    4
  MPG     {q2}, 300
  MD      {q2}
(repeat the previous 7 instructions once again)

  QNopReg r15        # Y180, Y180
  Pulse   {q2}, Y180
  Wait    4
  Pulse   {q2}, Y180
  Wait    4
  MPG     {q2}, 300
  MD      {q2}
(repeat the previous 7 instructions once again)

  ...

  addi    r1, r1, 1
  bne     r1, r2, Outer_Loop
\end{lstlisting}
\caption{QuMIS Program to perform \textit{AllXY} experiment.}
\label{alg:allxy}
\end{algorithm}

We have performed various quantum experiments on a qubit to validate and verify the design of QuMA and QuMIS, including $T_1$, $T_2$ Ramsey, $T_2$ Echo, \textit{AllXY}, and randomized benchmarking~\cite{epstein2014investigating} experiments. Considering the readability and page limitation, we only show the \textit{AllXY} experiment in the paper.

Figure \ref{fig:setup} shows the experimental setup. All classical electronics are at room temperature. The quantum chip, operating at 20 mK, contains 10 transmon qubits with dedicated readout resonators all coupled to a common feedline. The measured qubit (labeled 2) has transition frequency $\fq = 6.466~\GHz$, and the coupled resonator has fundamental $\fr = 6.850~\GHz$ (for qubit in $\ket{0}$) (further detailed in~\cite{bultink2016active}). To perform single-qubit gates, we use one microwave source [Rohde \& Schwarz (R\&S) SGS100A] to generate a 6.516 GHz carrier and control box AWG 2 to produce the in-phase and quadrature components (including $-50~\MHz$ single-sideband modulation) that define the pulse envelope. To generate the measurement pulse, we trigger a $6.849~\GHz$ carrier (generated by a R\&S SMB100A) using the control box digital output 1. The transmitted feedline signal is demodulated to an intermediate frequency of $40~\MHz$ using a $6.809~\GHz$ local oscillator (another R\&S SGS100A). Prior to the experiment, the qubit pulses are calibrated and uploaded into control box AWG 2. Since the operations in the \textit{AllXY} experiment are primitive, the micro-operation unit simply forwards the codewords to the wave memory without translation.

\begin{figure}[htp]
\centering
\includegraphics[width=0.9\linewidth]{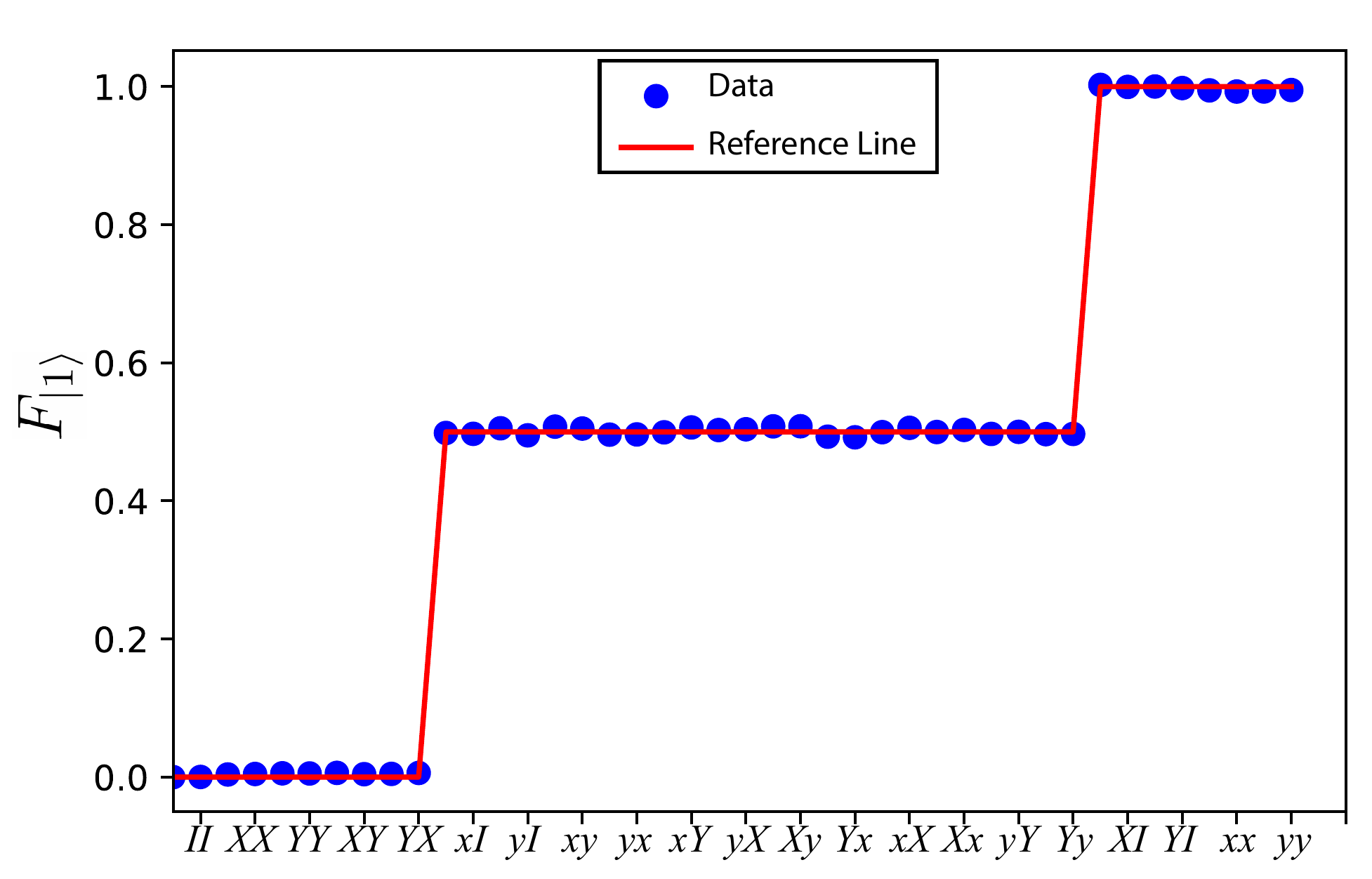}
\caption{The \textit{AllXY} result of qubit 2. In the label, each $X$/$Y$ ($x$/$y$) denotes a rotation by $\pi$ ($\pi/2$) around the $x$/$y$ axis of the Bloch sphere.}
\label{fig:allxy_result}
\end{figure}

The QuMIS program used to perform the \textit{AllXY} experiment is generated from a OpenQL description and is shown in Algorithm \ref{alg:allxy}. In this experiment, each of the 21 combinations is measured twice to make a direct visual distinction between systematic errors and low signal-to-noise ratio. Figure \ref{fig:allxy_result} shows the measurement results. The red staircase shows the ideal signature of perfect pulsing.
The results of the 0-th (18-th and 19-th) combination are taken as the calibration point $\bar{S}_{\ket{0}, r}$ ($\bar{S}_{\ket{1}, r}$). Using the calibration points to rescale the signal, we obtain the fidelity $F_{\ket{1}}|_i$ corrected for readout error:
\begin{align*}
    F_{\ket{1}}|_{\mathrm{meas},i} = \left(\bar{S}_i - \bar{S}_{\ket{0}, r}\right)/\left(\bar{S}_{\ket{1}, r} - \bar{S}_{\ket{0}, r}\right).
\end{align*}
We loop over these $K=42$ pulse combinations over $N=25600$ rounds. The data acquisition unit performs the required averaging of measurement results for each $K$.

This experiment uses the instructions generated from the high-level language OpenQL description to control the operations on the qubit. Only 7 pulses including the \textit{Identity} operation are stored in the lookup table of the codeword-triggered pulse generation unit, regardless of the number of combinations of operations. It has a moderate memory consumption to store $140~\ns\times R_s$ samples exhibiting a better scalability compared to the conventional method. From the experiment result, we can see that the measured fidelity for each combination matches well with the ideal readout fidelity. Since the \textit{AllXY} experiment is sensitive to imperfection of the pulses and the timing, it demonstrates that the right pulses are generated and the precise timing of operations is well preserved.

\section{Conclusion}
\label{sec:conc}
We have proposed and developed QuMA, a microarchitecture that takes the compiler generated instructions as input to flexibly control a superconducting quantum processor. Three mechanisms are introduced in QuMA to enable flexible control over quantum processors : i) codeword-based event control, ii) precise queue-based event timing control, and iii) multilevel instruction decoding pulse control mechanism. We have also designed and implemented the quantum microinstructions set QuMIS which can well describe quantum operations on qubits with precise timing.

We implemented a QuMA processor prototype on a FPGA. We have validated this microarchitecture by performing a successful \textit{AllXY} experiment on a superconducting qubit, using a combination of the auxiliary classical instructions and QuMIS instructions which are generated by OpenQL. QuMA enables flexible definition of quantum experiments by a straightforward change in the input program.

Future work will involve implementing a QuMA supporting a VLIW instruction set, and extending the microcode unit to enable the definition of quantum instructions and the execution of real-time feedback control.

\begin{acks}
We thank M.~Tiggelman, S.~Visser, J.~Somers, L.~Riesebos, E.~Garrido~Barrab\'{e}s, and E.~Charbon for contributions to an early version of the CBox, A.~Bruno for fabricating the quantum chip, H.~Homulle for drawing Figure~\ref{fig:stack}, and L.~Lao, H.~A.~Du~Nguyen, R.~Versluis and F.~T.~Chong for discussions. We acknowledge funding from the China Scholarship Council (X.~Fu), Intel Corporation, an ERC Synergy Grant, and the Office of the Director of National Intelligence (ODNI), Intelligence Advanced Research Projects Activity (IARPA), via the U.S. Army Research Office grant W911NF-16-1-0071. The views and conclusions contained herein are those of the authors and should not be interpreted as necessarily representing the official policies or endorsements, either expressed or implied, of the ODNI, IARPA, or the U.S. Government. The U.S. Government is authorized to reproduce and distribute reprints for Governmental purposes notwithstanding any copyright annotation thereon.
\end{acks}

\bibliographystyle{ieeetran}
\bibliography{reference.bib}

\end{document}